\documentclass[iop]{emulateapj}
\usepackage{natbib}
\usepackage{graphicx}
\usepackage{times}
\usepackage{xspace} 
\usepackage{booktabs} 
\usepackage{amsmath}
\usepackage{xcolor}

\slugcomment{\it Re-Submitted Version --- 2015 October 16}
\shorttitle{Infrared Time Lags for PG~1302-102}
\shortauthors{Jun et al.}

\def\akari{{\it Akari}}
\def\wise{{\it WISE}}

\def\pg{PG~1302-102}

\def\deg{\ifmmode {^{\circ}}\else {$^\circ$}\fi}
\def\kms{\ifmmode {\rm\,km\,s^{-1}}\else
    ${\rm\,km\,s^{-1}}$\fi}
\def\ergcm2s{\ifmmode {\rm\,erg\,cm^{-2}\,s^{-1}}\else
    ${\rm\,erg\,cm^{-2}\,s^{-1}}$\fi}
\def\ergAcm2s{\ifmmode {\rm\,erg\,cm^{-2}\,s^{-1}\,\AA^{-1}}\else
    ${\rm\,erg\,cm^{-2}\,s^{-1}\,\AA^{-1}}$\fi}
\def\ergs{\ifmmode {\rm\,erg\,s^{-1}}\else
    ${\rm\,erg\,s^{-1}}$\fi}
\def\kmsMpc{\ifmmode {\rm\,km\,s^{-1}\,Mpc^{-1}}\else
    ${\rm\,km\,s^{-1}\,Mpc^{-1}}$\fi}

\def\spose#1{\hbox to 0pt{#1\hss}}
\def\simlt{\mathrel{\spose{\lower 3pt\hbox{$\mathchar"218$}}
     \raise 2.0pt\hbox{$\mathchar"13C$}}}
\def\simgt{\mathrel{\spose{\lower 3pt\hbox{$\mathchar"218$}}
     \raise 2.0pt\hbox{$\mathchar"13E$}}}

\begin{document}

\title{Infrared Time Lags for the Periodic Quasar PG 1302-102}

\author{Hyunsung~D.~Jun\altaffilmark{1,2},
Daniel~Stern\altaffilmark{1},
Matthew~J.~Graham\altaffilmark{3},
S.~G.~Djorgovski\altaffilmark{3},
Amy~Mainzer\altaffilmark{1},
Roc~M.~Cutri\altaffilmark{4},
Andrew~J.~Drake\altaffilmark{3} \&
Ashish~A.~Mahabal\altaffilmark{3}
}

\altaffiltext{1}{Jet Propulsion Laboratory, California Institute
of Technology, 4800 Oak Grove Drive, Pasadena, CA 91109, USA [e-mail:
{\tt hyunsung.jun@jpl.nasa.gov}]}

\altaffiltext{2}{NASA Postdoctoral Program Fellow}

\altaffiltext{3}{California Institute of Technology, 1216 E.
California Blvd., Pasadena, CA 91125, USA}

\altaffiltext{4}{Infrared Processing and Analysis Center, California
Institute of Technology, Pasadena, CA 91125, USA}

\begin{abstract} 

The optical light curve of the quasar \pg\ at $z = 0.278$ shows a
strong, smooth 5.2~yr periodic signal, detectable over a period of
$\sim 20$~yr.  Although the interpretation of this phenomenon is
still uncertain, the most plausible mechanisms involve a binary
system of two supermassive black holes with a subparsec separation.
At this close separation, the nuclear black holes in \pg\ will likely merge within $\sim
10^{5}$~yr due to gravitational wave emission alone.  Here we report
the rest-frame near-infrared time lags for \pg.  Compiling data
from \wise\ and \akari, we confirm that the periodic behavior
reported in the optical light curve from \citet{Gra15a} is reproduced
at infrared wavelengths, with best-fit observed-frame 3.4 and
$4.6 \mu$m time lags of $(2219 \pm 153, 2408 \pm 148)$~days for a
near face-on orientation of the torus, or $(4103\pm 153, 4292 \pm
148)$~days for an inclined system with relativistic Doppler boosting
in effect.  The periodicity in the infrared light curves and the
light-travel time of the accretion disk photons to reach the dust
glowing regions support that a source within the accretion disk is
responsible for the optical variability of \pg, echoed at the further
out dusty regions.  The implied distance of this dusty, assumed
toroidal region is $\sim$ 1.5~pc for a near
face-on geometry, or $\sim$1.1~pc for the relativistic Doppler
boosted case.

\end{abstract}


\section{Introduction}

Optical reverberation mapping, measuring the time delay between
variations in the accretion disk optical continuum luminosity of
an active galactic nucleus (AGN) and the delayed response of the
broad emission line luminosity, has become a standard technique to
study the physics and structure of the inner regions of AGN (e.g.,
\citealt{Bla82}; \citealt{Pet93}; \citealt{Ben06}).  Such
studies enable measurement of the physical size of the broad line
region, opening a way to estimate the virial black hole masses of
distant AGNs (e.g., \citealt{Mcl02}; \citealt{Ves06}; \citealt{Jun15})
through the relation between broad line region size and optical
luminosity ($R_{\rm BLR} \sim L^{0.5}$; e.g., \citealt{Kas00}; \citealt{Ben06}).

Analogous to optical broad line reverberation analyses, light travel
time lags between the ultraviolet/optical and near-infrared continuum
provide a measurement of the physical size of the infrared-emitting
dusty tori that are believed to surround most AGN (e.g., \citealt{Pen74};
\citealt{Cla89}; \citealt{Sug06}). The inferred time-lag-based sizes
from this dust reverberation, supported by direct interferometric
$K$-band measurements for a few nearby AGN (e.g., \citealt{Swa03};
\citealt{Kis11a}), shows that the dust emitting region scales
with the ultraviolet/optical luminosity ($R_{\rm dust}-L$ relation;
e.g., \citealt{Okn01}; \citealt{Kos14}) with a similar
square root power as the $R_{\rm BLR}-L$ relation, but with a
larger normalization.  These radius-luminosity
relations indicate that the geometric components of an AGN are
spatially structured, such that illumination of the accretion disk
propagates first through the broad line and later through the cooler,
dusty regions.

Recently, \citet{Gra15a} reported an unusual example of an
optically variable quasar, \pg\ at $z =0.278$.  The 9-yr optical 
light curve from the Catalina Real-time Transient Survey (CRTS; \citealt{Dra09}), 
combined with historical data going back $\sim 20$~yrs, shows periodic variability 
with an observed period of $1884 \pm 88$~days.  Additional
examples of similarly periodic quasars have recently been reported
by \citet{Gra15b} and \citet{Liu15}.  This unusual
phenomenology is distinct from the typical variability of most
quasars which is akin to a damped random walk (e.g., \citealt{Mac12};
\citealt{Gra14}).  The most likely interpretations of the
periodic quasars involve close, subparsec binary supermassive
black hole systems \citep[e.g.,][]{DOr15a, DOr15b, Gra15b}.

\begin{figure*}
\centering
\includegraphics[scale=0.95]{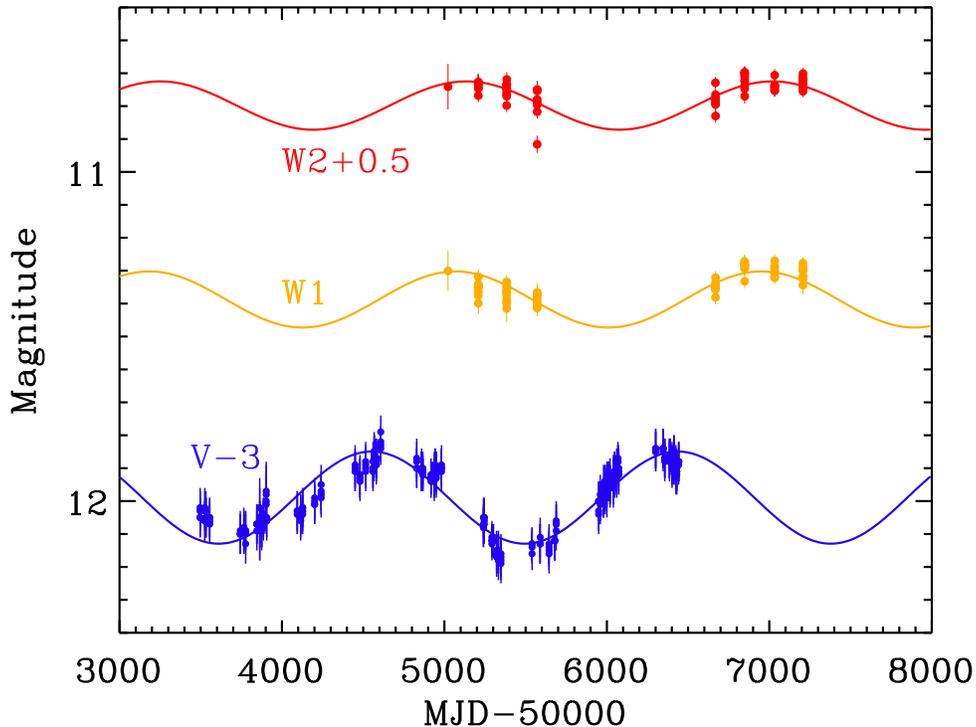}
\caption{The light curves for \pg\ in CRTS optical (blue), $\it WISE/NEOWISE$-R
$W1$ (yellow), and $W2$ (red). The earliest $W1$ and $W2$ magnitudes
are derived from the \akari\ $2.5-5.0 \mu$m grism spectra (\S 2). Overplotted
sinusoids are the best-fit curves from \citet{Gra15a} and this work (\S 3), in the
same colors as each data.} 
\end{figure*}

Taking advantage of the distinct periodic signals in the optical,
we investigate the infrared light curves of \pg\ using data from
the {\it Wide-field Infrared Survey Explorer} ({\wise; \citealt{Wri10})
and \akari\ \citep{Mur07} missions.  We also include more recent
data from the {\em Near-Earth Object} \wise\ {\em Reactivation} mission ({\it
NEOWISE-}R; \citealt{Mai14}).  We measure the time lags between
the optical and infrared emission, thereby confirming that the
optical periodicity is reproduced in the dust emission and provides
a size measurement of the dust-emitting region. Furthermore, using
multi-band $3.4 \mu$m ($W1$) and $4.6 \mu$m ($W2$) data, we 
investigate wavelength-dependent trends in the light curve time
delays, which provide further information about the size and structure
of the dusty torus.  In a separate, related paper, \citet{DOr15b}
report on the UV light curve of \pg, showing that it is consistent
with relativistic boosting from a supermassive black hole binary being
the cause of the observed UV and optical periodicity.

In this {\it Letter}, we describe the data collection and processing
in \S 2, the measurement of time-lags and amplitudes in \S 3, and
their implication for the binary supermassive black hole interpretation
in \S 4.  Throughout, photometry is reported in the Vega system and
we adopt a flat $\Lambda$CDM cosmology with $H_0 = 70\ \kmsMpc$,
$\Omega_m =0.3$ and $\Omega_\Lambda =0.7$.

\section{Mid-Infrared Observations}

\pg\ is bright enough to have been detected in single exposures
throughout the \wise\ mission. Collecting both cryogenic and
post-cryogenic multi-epoch photometry from the AllWISE and {\it
NEOWISE-}R data releases, as well as the most recent (pre-release)
{\it NEOWISE-}R observations of \pg, we measure $W1$ and $W2$ light
curves spanning 5.5~years (MJD 55208-57207). The data consist of
seven groups spaced at intervals of roughly six months, albeit with
a three year gap when \wise\ was in hibernation (MJD 55572-56668).
Each group has 10-15 observations, with the latest data taken in
July 2015. {\it NEOWISE-}R is expected to obtain $\sim 3$ additional
groups of observations over the next $\sim1.5$~yrs.  We do not
include longer wavelength data at $12 \mu$m ($W3$) or $22 \mu$m
($W4$) which are restricted to the cryogenic period of the \wise\
mission as the temporal coverage of those observations consist of
only two groups. We further removed the data suffering
from bad quality frames ({\it qual\_frame=0}), charged particle
hits ({\it saa\_sep$<$0}), or scattered moonlight ({\it moon\_masked=1}).

Supplementing the \wise\ observations of \pg, we also analyzed archival
$2.5-5.0 \mu$m grism spectra obtained by \akari\ six months prior to the
\wise\ observations \citep{Ohy07}.  Out of three post-cryogenic \akari\ spectra
of \pg\ taken on MJD 55022-55023, we removed the second spectrum as it is
significantly affected by hot pixels.  We convolved the processed
and extracted spectra from the other two observations with the
\wise\ filter responses \citep{Wri10} to obtain $W1$
and $W2$ equivalent magnitudes. Because the $W2$ filter response
extends longward of $5 \mu$m, the wavelength cutoff of post-cryogenic
\akari, we extrapolated the spectrum redward based on a power law
fit, taking into account the $1 \sigma$ uncertainties in the fitted
parameters.  Finally, we averaged the two \akari\ magnitudes, separated by one day, to improve the sensitivity.
For ease of description, in the following we discuss the $W1$ and
$W2$ light curves, though those light curves also include these
earlier epoch \akari\ observations. Figure~1 presents the CRTS, \wise, and \akari\
light curves with the periodic functional fits (\S 3) overplotted.

\begin{figure*}
\centering
\includegraphics[scale=0.97]{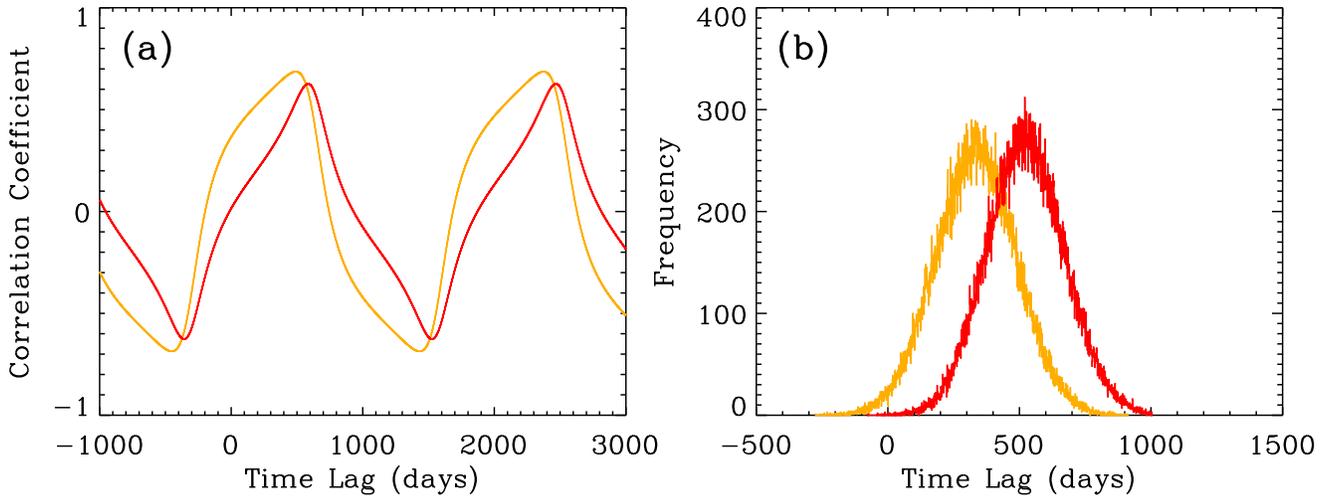}
\caption{(a) The cross correlation amplitudes for \pg\ between the best-fit optical model and the \wise\
$W1$ (yellow) and $W2$ (red). (b) Monte Carlo simulation of the time lags for $W1$ (yellow) and $W2$ (red) 
taking into account the uncertainties in the observed magnitudes and period.}
\end{figure*}

\section{Measurement of the Time Lags}
Assuming that the optical variability of \pg\ arises from the
accretion disk(s) of the (binary) AGN, we expect the fluctuations in the optical
luminosity to process through the inner dusty torus and produce a distribution
of time lags in the hot dust emission corresponding to the light travel time to the
inner portions of the torus. We aim to measure the cross-correlation signal  
between the optical and infrared magnitudes, and test if the infrared light curves 
trace the periodic variability appearing in the optical time series data.

Because the optical light curves in \citet{Gra15a} show a well defined 
periodic behavior, we first cross correlate their sinusoidal model fit with the 
$W1$ and $W2$ band data. In Figure 2(a) we plot the cross correlation amplitudes 
for a range of time lags. The cross correlation amplitudes peak at 0.69 for $W1$ 
and 0.63 for $W2$, indicating that the \wise\ data are moderately correlated to the 
periodic optical light curve model. 

To give proper treatment of the uncertainties 
into the cross correlation analysis (e.g., \citealt{Pet98}), we performed a set of 
Monte Carlo simulations including the observational uncertainty from the CRTS data (0.06~mag) 
and the uncertainty in the optical period (88~days). 
We performed 100,000 realizations adding random Gaussians to the 
model magnitudes with $\sigma=0.06$~mag, and modifying the period 
by adding random Gaussians with $\sigma=88$~days. We measured the centroid of 
each simulated cross correlation function where the amplitudes were above 0.8 times 
the peak value, while limiting the range of time lags to $[-500,1500]$~days to do away 
with the other periodic solutions. In Figure 2(b) we plot the resultant distribution of the 
time lags from the Monte Carlo simulations. Adopting the fiducial time lag as the mean of the 
distribution of simulated time lags, and the uncertainty as the 34.13\% percentile 
in the time lag distribution deviated from the mean \citep{Pet98}, the $W1$ and $W2$ 
time lags to the optical model ($\tau$, observed-frame days) become 
\begin{eqnarray}\begin{aligned}
\tau_{W1}= (335 \pm 153)+1884n\\
\tau_{W2}= (524 \pm 148)+1884n
\end{aligned}\end{eqnarray}
with any integer ($n = 0, \pm 1, \pm 2, ...$) multiples of 1884~days
providing equally probable solutions (at least mathematically, if
not physically).

The sizable cross correlation amplitudes to the periodic model imply that 
the \wise\ light curves are also periodically variable. We checked this by 
fitting the infrared light curves of \pg\ with periodic functions and calculating the signal-to-noise (S/N)
of the periodic modulation above white noise. Adopting the sinusoidal period from \citet{Gra15a},
we fit the magnitude $m(t)$ at the observed-frame time $t$ as $m(t)=A \cos (2\pi t/T)+B$,
solving for amplitude $A$ and magnitude zero-point $B$, and fixing the period $T=1884$ days. 
Applying the best-fit values into the periodic S/N ratio indicator of \citet{Hor86}, 
we get values of 5.7, 4.1, and 3.4 in the optical, $W1$, and $W2$,
respectively. The false alarm probability $F$ that a periodicity in the data 
arises from pure noise, equations~(13) and (23) from \citet{Hor86}, yield $F \sim 3 \times 10^{-50}$ 
from the CRTS data alone, and $\sim 4 \times 10^{-10}$ and $\sim 1 \times 10^{-9}$ for $W1$ and $W2$, 
where these estimates of $F$ were calculated assuming evenly spaced data. 
The extremely small $F$ values imply with high likelihood that the source is periodic in all three wavebands.

We next attempt to estimate the most reasonable time lag between
the optical and mid-infrared light curves, $\tau$ in equation~(1),
out of the infinite number of solutions allowed.
First, we assume that the time lags are positive such that they
represent the reverberation of the incident optical photons through
dust re-processed emission, requiring $n \geq 0$ in equation~(1). 

Further refinement of the value of $n$ requires a physical model of
the system.  In the general picture of a toroidally shaped dusty torus, 
most of the mid-infrared emission at these wavelengths will
come from the inner edge of the torus, with longer wavelength
emission coming from slightly larger radii, corresponding
to cooler material (e.g., \citealt{Nen08}).
For an axisymmetric model of the torus simply responding to variations in the accretion
disk luminosity, the time delay then depends on the light travel time
to the inner edge of the torus and the orientation of the torus relative to our line of sight, 
$\tau = R\, (1+\sin i)\, /c$, where $R$ is the radius of the inner edge of the torus, 
$c$ is the speed of light, and $i$ is the inclination angle ($i=0\deg$ corresponds to face-on; $i=90\deg$ corresponds to edge-on). 
Alternatively, \citet{DOr15b} present a compelling and physically motivated
model for \pg\ where the modulated UV/optical emission is due to relativisic Doppler boosting
of emission from a ``minidisk'' in a steadily accreting, unequal mass binary.  In 2D and 3D
hydrodynamical simulations of supermassive black hole binaries, circumprimary and circumsecondary
minidisk accretion flows form around the two black holes, with a larger circumbinary accretion
disk on larger scales (e.g., \citealt{Shi12}; \citealt{DOr13}).  For an unequal mass system, the secondary 
carves out an annular region in the circumbinary disk, limiting the amount of material that
reaches the primary.  Most of the emission at optical through X-ray energies will thus be associated
with the rapidly accreting secondary, which \citet{DOr15b} show is likely moving at $\sim 0.07c$ with an 
orbital inclination of $i=60-90\deg$ for the \pg\ system.  
Relativistic Doppler boosting then produces a system akin to a rotating lighthouse, producing
a sinusoidal light curve with a period $T$.  Whereas the peaks of the UV and optical light occur
when the ``lighthouse'' is pointed towards the observers line-of-sight, peaks in the thermal mid-infrared
emission, which is predominantly coming from the inner wall of the far side of the torus,
correspond to when the ``lighthouse'' is pointed away from the observer, giving an additional 
$T/2$ to the infrared time lag.  

We can now use these models to place limits on the dust lags based
on the sublimation radius of hot dust. The physically motivated
lighthouse model is effective when the inclination is large so that
the relativistic boosting is working, and the near side of the torus
is blocked from our line of sight.  Thus, our variable circumbinary
accretion disk model predicts $\tau = R\, (1+\sin i)\, /c$ and is
appropriate for small $i$, while the relativistic Doppler boosting
model predicts $\tau = R\, (1+\sin i)\, /c + T/2$ and requires large
$i$.  
The inner radius of
the torus corresponds to the dust sublimation radius $R_{\rm sub}$,
i.e., the closest distance to the central engine that dust is not
destroyed, and we solve for $n$ in the $W1$ time
lag from equation~(1) where the two radii are closest to each other,
or $R \simeq R_{\rm sub}$.  Graphite dust grains are better able
to survive in the hotter, inner region of the torus than silicate,
and they are thought to be responsible for the rest-frame near-infrared
emission from quasars (e.g., \citealt{Mor12}; \citealt{Jun13}).  We
use the graphite dust sublimation radius $R_{\rm sub} \simeq 0.5
(L_{\rm bol}/10^{46}\,{\rm erg}\, {\rm s}^{-1})^{0.5}$~pc for a
sublimation temperature $T_{\rm sub} =1800$~K \citep{Mor12}.  At
$z = 0.278$ and a median $V \simeq R\simeq 15.0$~mag for \pg\
(\citealt{Mar96}; \citealt{Ojh09}; \citealt{Pro11}; \citealt{Gra15a}),
$L_{\rm bol} \simeq 6.78 \times 10^{46}\, {\rm erg}\, {\rm s}^{-1}$
adopting a 5100\AA\ to bolometric correction of 10.33 \citep{Ric06}.
Thus, we have $R_{\rm sub} \simeq 1550$~rest-frame light days
(1.30~pc), or $R_{\rm sub} \simeq 1980$~observed-frame light days.
For a near face-on orientation of the torus ($i \simeq 0 \deg$), $\tau =
R\, (1+\sin i)\, /c \simeq R\, /c$ and the $n=1$ solution from
equation~(1), ($\tau_{W1}, \tau_{W2}) = (2219\pm 153, 2408 \pm
148)$~days, best meets $R \simeq R_{\rm sub}$ from the $W1$ time
lag, with $R=1.12R_{\rm sub}$=1.46~pc.  Meanwhile, for a relativistic
Doppler boosted torus with $i=60-90 \deg$, the $n=2$ solution,
($\tau_{W1}, \tau_{W2}) = (4103\pm 153, 4292 \pm 148)$~days,
gives the best match between $R$ and $R_{\rm sub}$ with $R=0.80-0.86R_{\rm
sub}=1.04-1.11$~pc.  We note that the uncertainty in the $R \simeq
R_{\rm sub}$ approximation could be a few tens of percent \citep{Mor12},
and that even with the same adopted $T_{\rm sub}=1800$~K, the value
of $R_{\rm sub}$ varies across the literature by 20-60\% due to
slightly different model assumptions (e.g., \citealt{Bar87};
\citealt{Nen08}). Also the uncertainty in the bolometric correction from the 
mean quasar template could be as large as 50\% \citep{Ric06}, translating 
into 25\% uncertainty in $R_{\rm sub}$.

We now consider the most probable \wise\ time lags 
in the context of the observed $R_{\rm dust}-L$ relation from \citet{Kos14}.
[Note that since \citet{Kos14} studied variable
but non-periodic AGN, nor is the Doppler-boosted lighthouse model likely
appropriate for their single black hole systems,
their work does not suffer the same degeneracy
with integer multiples of the variability period.] The \wise\ $W1$
and $W2$ filters correspond to rest-frame 2.6 and $3.6 \mu$m at $z
= 0.278$.   Using the median optical magnitudes of $V \simeq R \simeq 15.0$, the
rest-frame $V$-band luminosity of \pg\ is $L_V \simeq 6.56 \times 10^{45}\,
{\rm erg}\, {\rm s}^{-1}$.  Based on the $K$-band $R_{\rm dust}-L$ relation \citep{Kos14}, 
the expected observed-frame time lag at $z = 0.278$ is $\tau \simeq 1590 \pm 150$~days. 
Our $W1$ and $W2$ lags are above the \citet{Kos14} relation
by 0.15 and 0.18 dex for $n=1$, and 0.41 and 0.43 dex for $n=2$.
The former offsets are comparable to the observed scatter of the relation (0.14\,dex), 
but the latter are not, consistent with the $T/2$ delay effect 
and the large inclination angle required by the lighthouse model. 
We also note that the $K$-band $R_{\rm dust}-L$ relation generally predicts $R_{\rm dust}$ 
smaller than $R_{\rm sub}$ by factors of a few, and \citet{Kos14} details how $R_{\rm sub}$ 
depends both on dust grain size and the details of the dust grain absorption efficiency. 
Also, \citet{Nen08} note that the largest grains could survive to closer radii than $R_{\rm sub}$, 
presumably detected by the $K$-band time lags.

Previous near-infrared studies of local AGN have shown
dust lags are typically longer at longer near-infrared wavelengths,
since cooler material (e.g., further out material) will dominate
the emission at progressively longer wavelengths (e.g., \citealt{Kis11b};
\citealt{Vaz15}). The $W2$ time lag is consistent with, or only marginally larger than the $W1$ lag 
with a ratio $1.09 \pm 0.10$ for the $n = 1$ and $1.05 \pm 0.05$ for the $n = 2$ solution in equation~(1).
Following \citet{Vaz15}, we quote the radial distribution of dust grains in 
radiative equilibrium (e.g., \citealt{Bar87}; \citealt{Nen08}) scaling with temperature as 
$R_{\rm dust} \sim T^{-2.6}$ and find $R_{{\rm dust,}W2} /R_{{\rm dust,}W1}
\simeq 2.3$, much higher than the value from this study. 
Alternative explanations such as clumpy tori \citep{Nen08} are
successful in emitting a broad infrared spectrum from a confined
distribution of dust.

\section{Discussion}

Graham et al. (2015ab) list several possible physical mechanisms
to produce the periodic optical light curves in \pg:  non-thermal
contribution from a precessing jet in a supermassive black hole
binary, periodic mass accretion in a circumbinary accretion disk,
or a precessing warped disk (which might also be related to a binary
supermassive black hole).  Our results suggest that the observed
mid-infrared variability is being driven by variations in the accretions disk(s). 
For example, in the precessing jet scenario, one
would expect very little temporal offsets, if any, between the
optical and infrared variability.  Indeed, the observed periodic
signal in the \wise\ data and the cross correlation results support that 
dust is reverberating input luminosity from the accretion disk.

The elliptical host galaxy morphology of \pg\ and the presence of
close companion galaxies around it (\citealt{Bah95}; \citealt{Dis95})
are suggestive of a recent merger driving high-luminosity
AGN activity (e.g., \citealt{Tre12}; \citealt{Hon15}).  Indeed,
the host galaxy of \pg\ measured from $H$- and $R$-band imaging are
classified as asymmetric and tidal tail morphologies, respectively
(\citealt{Guy06}; \citealt{Hon15}). The galaxy-scale morphology hints that this system may be in an ongoing merger.

The mid-infrared light curves of \pg\ are indicative of periodic
dust reverberation lagging behind the optically periodic light curve
due to the light travel time to the mid-infrared emitting dusty 
torus, with a possible additional phase lag in the relativistic 
Doppler boosting lighthouse model. For both models, the time lags 
imply that the dust is echoing outside the accretion disk but from
a small, possibly clumpy medium, and we predict that the narrow flourescent 
Fe K$\alpha$ X-ray emission should vary on the same period and 
with the same phase lag as the mid-infrared emission.
Time domain studies are starting
to reveal events in AGNs occurring on physical scales that
are not resolvable from the Earth, helping to constrain the source
of variability and properties of the variable structure (e.g., \citealt{Ris13};
\citealt{LaM15}; \citealt{Meh15}; \citealt{Ste15}). 
In particular,
multiwavelength time domain surveys of AGN combining optical data
with mid-infrared data from \wise, as done here, or from the {\it Spitzer
Space Telescope} (e.g., \citealt{Gor14}; \citealt{Vaz15}), probe the
physical structure and accretion mechanisms of the AGN accretion
disk and surrounding dusty material.

\acknowledgements 
We thank the anonymous referee for the comments which greatly
improved the paper, as well as Daniel D'Orazio, Moshe Elitzur, 
Saavik Ford, Zoltan Haiman, Barry McKernan, and Robert Nikutta 
for helpful discussions.
This research was supported by an appointment to the NASA Postdoctoral
Program at the Jet Propulsion Laboratory, administered by Oak Ridge
Associated Universities through a contract with NASA. This publication
makes use of data products from the {\it Wide-field Infrared Survey
Explorer}, which is a joint project of the University of California,
Los Angeles, and the Jet Propulsion Laboratory/California Institute
of Technology, funded by the National Aeronautics and Space
Administration. This publication makes use of data products from
{\it NEOWISE}, which is a project of the Jet Propulsion
Laboratory/California Institute of Technology.  {\it NEOWISE} is
funded by the National Aeronautics and Space Administration.  CRTS
was supported by the NSF grants AST-1313422 and AST-1413600.  DS
acknowledges support from NASA through ADAP award 12-ADAP12-0109.

\smallskip
{\it Facilities:} \facility{CRTS}, \facility{WISE}, \facility{NEOWISE}

\smallskip
\copyright 2015.  All rights reserved.

\clearpage
\end{document}